\documentclass{llncs}
\usepackage[english]{babel}
\usepackage{amsfonts}
\usepackage{amssymb}

\usepackage[cmex10]{amsmath}

\usepackage{ifpdf}

\usepackage{cite}

  \usepackage[pdftex]{graphicx}
  \graphicspath{{../pdf/}{../jpeg/}{./BRS/}}
\PassOptionsToPackage{hyphens}{url}
\usepackage[pdftex,
            pagebackref=true,
            colorlinks=true,
            linkcolor=blue
           ]{hyperref}

\usepackage[tight,footnotesize]{subfigure}
\def\email#1{\href{mailto://#1}{#1}}

\begin{document}
\renewenvironment{quote}
               {\list{}{\rightmargin\leftmargin}%
                \item\relax\small\flqq\ignorespaces}
               {\unskip\unskip\frqq\endlist}
%

\title{How Resilient Are Our Societies?\\
	Analyses, Models, and Preliminary Results}
       \author{Vincenzo De Florio and Arianit Pajaziti}
       \institute{MOSAIC/University of Antwerp and MOSAIC/iMinds research institute\\
Middelheimlaan 1, 2020 Antwerp, Belgium\\
\email{vincenzo.deflorio@gmail.com}, \email{arianitpajaziti@gmail.com}}

\maketitle

\begin{abstract}
Traditional social organizations such as those for
the management of healthcare and civil defence
are the result of designs and realizations that matched well with
an operational context considerably different from the one we are experiencing today:
A simpler world, characterized by a greater amount of resources to match
less users producing lower peaks of requests.
The new context reveals all the fragility of our societies:
unmanageability is just around the corner unless we do not
complement the ``old recipes'' with smarter forms
of social organization. Here we analyze this problem and
propose a refinement to our fractal social organizations as
a model for resilient cyber-physical societies.
Evidence to our claims is provided by simulating our model
in terms of multi-agent systems.

\end{abstract}


%

\def\PS{\mathcal{P\kern-.2ex S}}
\def\CONTEXT{\hbox{$\mathcal{C}$}}
\def\C{\hbox{$\mathcal{C}$}}
\def\PREC{\prec_{\mathcal{P}}}
\def\PRECA{\prec_{\mathcal{A}}}
\def\PRECMA{\prec_{\mu\kern-.3ex\mathcal{A}}}
\def\PRECMC{\prec_{\mu\kern-.2ex\mathcal{C}}}
\def\PRECMR{\prec_{\mu\kern-.2ex\mathcal{R}}}


\section{Introduction}\label{s:intro}

Regardless of its nature, any system is affected by its design assumptions. Our societies are no exception.
The emergence of sought properties such as economic and social welfare for all;
sustainability with respect to natural ecosystems;
and especially manageability and resilience, highly depends on the way social organizations are designed.
A typical case in point is given by traditional organizations
operating in domains such as healthcare and crisis management.
A common assumption characterizing those organizations is the adoption of
a strict client-server model. This produces at least the following major consequences:
\begin{description}
\item[Stigmatization.]
   Users are permanently classified into service providers 
   and service receivers. Traditional healthcare organizations
   for instance typically classify users into two disjoint categories: the
   active users, namely
   professional and informal carers, and the patients and the elderly, who are
   considered as users incapable of any active behavior~\cite{SDGB10b}.
   Likewise, in
   disaster management organizations, predefined active roles are assigned to institutional responders
   while the citizens are confined to a passive role~\cite{CARRI3,DFSB14}.
\item[Fragility.]
   The artificial distinction in an active and a passive side of society severely affects
   quality-of-emergence~\cite{DF15a,DFSB14}.
   In particular, it
   introduces a systemic performance penalty in that only a subset of the social actors is available
   to serve the whole set.
   As well known, the fast growth and the progressive aging of the human population are introducing
   a new context---one in which the ``servicing subset'' is quickly decreasing in proportion.
   The problem is no more the ever increasing social costs; rather, it is the fact that
   the spectre of unmanageability---namely the vision of
   a fragile society unable to serve its citizens---is just around the corner.
\item[Absence of a referral service.]
   Despite having an only partial view on the capability and current state of the available servers,
   it is the responsibility of the client to identify which server to bind to. 
   It is the user that needs to know, e.g., which emergency service to invoke; which
   hospital to call first; which civil organization to refer to, and so on.
   Referral services \emph{do\/} exist, though they mainly cover a single domain (i.e., healthcare)
   and very specific and simple cases (typically, the seamless transfer of patient information
   from a primary to a secondary practitioner~\cite{EReferrals}).
   Because of this specialization
   such services mostly possess an incomplete view of the available resources.
\item[Lack of unitary responses to complex requests.]
   To the best of our knowledge, none of the existing referral services provides a composite response to complex
   requests such that the action, knowledge, and assets
   of multiple servers are automatically or semi-automatically combined and orchestrated.
   Even electronic referral systems in use today are mostly limited~\cite{ShBe07} and only provide
   predefined services in specific domains\footnote{%
	Interesting examples of such systems include SHINE~\cite{SHINE} and SHINE OS$+$~\cite{SHINEOS+}.}.
   As a consequence, 
   in the face of complex servicing requests calling for the joint action of multiple servers,
   the client is basically \emph{left on its own}. Societal organizations do not
   provide unitary responses nor assist the client in composing and managing them.
   Reasons for this may be found in lack of awareness and also in the ``convenient''
   shift of responsibility for failures 
   from the server to the client\footnote{As observed by~\cite{ZenVerhulst},
	an example of said shift of responsibility may be found in car industry with respect to
	aviation industry. A matter for reflection is the fact that the shift of responsibility
	regrettably translates in an inferior safety culture.}.
\end{description}


A logic consequence of the above situation is the urgent need to mutate our organizational paradigms
and assumptions. Simply stated, we cannot afford anymore not to use the full potential of our societies.
This means that the artificial distinction between an active and a passive subset should be removed---or
at least significantly reduced.
Moreover, the increasing complexity of modern times require that
\emph{societal organizations assume responsibility for becoming the enablers of collectively intelligent responses}.
New organizational design assumptions are called for,
able to provide us with new servicing paradigms---in other words, new ways to perceive
and manage the status quo.
The vision of the organization as a system restricting the freedom to play roles
should be changed into that of an enabler and a provider. Instead of preventing
participation, the organization should allow roles to be filled in by whomever is able
and willing to participate. More than this,
\emph{the organization should function as a catalyst of mutualistic cooperation\/}
among the role players at all levels,
from the citizens to the governing institutions. By means of the organization,
knowledge should flow among the players highlighting needs, assets, requirements, and
opportunities. The organization should assist in the process of self-orchestrating
a response, making it easier for all parties involved to coordinate themselves, exchange
information, and take the right and timely decisions.

Two key challenges of our societies are, one the one hand, being able to
define one such organizational model.
At the same time, fundamental aspects of the \emph{identity\/} of the organization must be preserved.
A second challenge is thus learning how to guarantee the resilience of our ``evolved'' organizations.
Organizational fidelity~\cite{DF14a,DBLP:journals/corr/FlorioP15} (guaranteed bounds to identity drifting)
therefore becomes a new design requirement for our future organizational models.

In this paper we present preliminary results obtained simulating an organizational model called
Fractal Social Organizations (FSO).
FSO takes the responsibility as
a generalized referral service for the orchestration
of complex social services. In FSO, the rigid client-server scheme of
traditional organizations is replaced by service orientation while bottleneck-prone hierarchies are
replaced by communities of peer-level members. 
Role appointment is not static and directed by the organization, but rather
voluntary and context-driven. It is our conjecture that the just stated 
new design assumptions allow for the creation of smarter societal organizations
able to match the complexity of our new complex world.

The structure of this article is as follows.
%
In Sect.~\ref{s:fso} we briefly recall the main assumptions of FSO.
Section~\ref{s:failures} reports about real-life facts
that highlight some of the deficiencies of traditional organizations.
Those facts inspired the first scenario introduced in
Sect.~\ref{s:scenarios}, which represents a case of inefficient interaction between
clients and servers of a traditional organization. In the same section we derive two
alternative scenarios based on FSO and highlight how those may result
in a more satisfactory interaction.
Section~\ref{s:model} introduces our simulation model.
%
Results are then reported in Sect.~\ref{s:results}.
Our preliminary conclusions and a view to future work are finally provided in Sect.~\ref{s:end}.
%

\section{Fractal Social Organizations}\label{s:fso}


This section provides a concise introduction to FSO. A more in-depth discussion is given in~\cite{DF13c,DFSB13a,DBLP:journals/corr/FlorioSB14}.

FSO replaces the classic client-server approach with service orientation and hierarchical organization with 
communities of users. Traditional assumptions are replaced by the following ``axioms'':

\begin{description}

\item[Agent.] An agent is an active-behaviored system, e.g. a person, an animal, a cyber-physical system, or a collection thereof.
  Agents are endowed with properties (for instance, mobility) and with assets (objects or agents.)

\item[Community.] 
  FSO may be described as a hierarchy of so-called Service-oriented Communities (SoC), namely 
  social organizations of agents in mutual (physical or logical) proximity. Communities are associated
  either to functional places---for instance a flat, a building~\cite{DFSB13a}, a hospital, or a region---or
  to loci---places characterized by the occurrence of some event, such as an accident or a fire. As a collection
  of agents, an SoC is also an agent, thus it can be part of a community. Conversely, an agent and its assets
  constitute a community. In other words, agents and communities are convenience aliases referring
  to the same concept.
  A community's parent community shall be referred in what follows as the higher-up.

\item[Community representative.]
  An SoC community is represented by one of its agents. Said representative agent may be the ``owner'' of the community---typically when the 
  community is an agent together with its assets---or it may be an agent with special features that is elected to play such role.
  In the first case the community is called an ``individual SoC'' (or iSoC for short).
  The community representative is the \emph{personification\/} of its SoC and has a 
  dual nature: it is a member of the community it represents and at the same time it is a member of the higher-up.

\item[Role.]
  Agents of an FSO can take a role, namely agree to accomplish actions belonging to a certain class of activities (for instance
  surgeon-, driver-, nurse-, or patient-specific activities).

\item[Notification.]
  Agents can publish notifications.
  Notifications can be status notifications, requests for service, notices of availability, etc. Notifications published by
  agent $a$ reach the representative of the community $a$ is a part of.

\item[Match.]
  For each new notification it receives, the community representative performs a semantic match: it verifies semantically~\cite{SDB13a}
  whether the new notification and those already known ``enable'' actions. 
  
\item[Action.] By ``action'' it is meant here some function of roles that becomes activated when 
  all the necessary ``ingredients'' (i.e., roles) are available. Actions are therefore similar to reservation stations
  in Tomasulo's algorithm for dynamic instruction scheduling in computer processors~\cite{Tom67}. An action whose roles are
  all available is called a resolved action.

\item[Exception.]
  If an action is semantically labeled as ``critical'', the lack of a role triggers an exception: the 
  community representative propagates the notification to the higher-up. This goes on until all roles are found
  in the traversed SoC's or until a flooding threshold is overcome and a failure is issued.
  
\item[Social Overlay Network.]
  When an action is enabled, the corresponding actors become a new temporary SoC whose lifespan in limited to the
  duration of the action.  
  If enabling an action required exceptions, the new temporary SoC is made of agents from different communities---in other
  words, the new community brings together nodes from different layers of the FSO hierarchy. Because of this we call
  such new SoC a Social Overlay Network (SON).
  
\item[Antifragility.]
  While actions are resolved, new SONs are assembled, launched, and finally dismissed. In the process, the FSO tracks the performance
  of the individual agents within a SON (the ``Parts'') as well as the performance of the overall SON (the ``Whole'').
  Said knowledge is used when assembling future SONs in a manner similar to that used in~\cite{Buys2011,BDFB12}.
  Machine learning may be used to reach antifragility~\cite{2015arXiv150308421D,Jones2014870,Taleb12,DF14a} and, e.g.,
  augment systematically the quality of the enrollment process as well as
  reapply the ``best'' schemes across the scales of the FSO.
\end{description}

From the above discussion we can characterize an
FSO as a social organization of SoCs---in other words, the ``root'' of a social hierarchy of communities.
By the above axioms of Agent and Community, FSO and SoC also are convenience aliases matching the same concept.

Working as a sort of ``cybernetic sociocracy''~\cite{DFSB13a}, FSO is based on inter- and intra-community 
cooperation and on the establishment of mutualistic relationships between members~\cite{DBLP:journals/corr/FlorioSB14,SDB13a}.
The rigidity of existing hierarchies is thus replaced with an agile and dynamic infrastructure that exploits cooperation across the scales 
of the service. This enables what we could refer as a new ``general mode of operation'' for social organization, which can be schematized
as follows:

\begin{itemize}
\item An event takes place and creates a new ``need''. As a result, an agent of an SoC issues a notification.
	
\item The notification reaches the SoC representative---for instance, a healthcare institution.
	
\item The community representative tries to manage the response locally. If an optimal response can be found within
  the community (or, in other words, all roles can be allocated in the current ``region''), the response is
  enacted and the state is adjusted. If no optimal response can be found,
  an exception propagates the need and the current response state to the higher-up. Both need and response
  ``travel'' through the levels of the FSO hierarchy until the response is finally enabled or a failure is declared.
	
\item While being executed, new knowledge is accrued both locally and globally. The current and future responses are thus refined,
  re-verified, and if (safer, cheaper, or better performing) alternatives are found, they are (and, from then on, will be) considered. 
\end{itemize}

%

\section{Failures}\label{s:failures}
In this section we report about a few real-life cases.
In all such cases the interactions between individuals and traditional organizations proved
to be ineffective, which led to severe consequences for the users.

\begin{itemize}
\item
    ``Newborn child dies in the ambulance before reaching a hospital''~\cite{FSOcase-Catania}.

\begin{quote} \ldots After a normal delivery, the newborn accused difficulty in breathing. The doctors tried in vain 
to locate a hospital where she could be transferred. The emergency service initiated a monitoring in the three 
hospitals in Catania in which pediatric intensive therapy is available: the Garibaldi, the Holy Child and 
Cannizzaro. None of the three centers had a free bed. The only hospital in eastern Sicily that responded to the 
call was Ragusa, more than a hundred miles away---a good hour travel. Sadly the little girl died during transportation in 
a private ambulance.
\end{quote}

\item ``Girl dies of common cold while searching for a hospital''~\cite{FSOcase-Milazzo}.

\begin{quote} \ldots A hospital in Milazzo (IT) did not have an Intensive Care Unit, and none was available in 
any of the other hospitals in the Milazzo province. Eventually, a place was found 200 miles away, in Caltagirone, 
on the other side of Etna. After two hours driving, the young girl died.
\end{quote}

\item ``52 years old man died yesterday aboard an ambulance, while looking for a hospital to have operations
for a double aneurysm''~\cite{FSOcase-Rome}.

\begin{quote}
\ldots The sick was transported to the hospital Umberto I. There, unfortunately, cardiac surgery was ``locked''
for technical issues since March 24. Rossella Moscatelli, deputy director of health at Umberto I
explains: ``For two weeks we have been informing all hospitals, including Colleferro and the emergency 
service, that our temperature controllers were out of order''. Those devices are essential to run the machinery 
for extracorporeal blood circulation during open heart surgery. Yesterday the new temperature controllers, just 
tested, ``were in operation for two planned by-pass surgeries''---she adds---``and Colleferro was informed
i\emph{only five minutes before the arrival of the ambulance}:
as they tried to locate in the city another hospital available, we 
attempted to stabilize the patient. \emph{After nearly two hours\/} we managed to find a hospital available, the San 
Camillo. For the patient, however, it was too late''. Simona Pasca of ``Cittadinanza Attiva---Tribunal for 
Patients' Rights'' remarks: ``\emph{If hospitals were better connected to each other},
that patient might be still alive. 
\emph{The problems of the\/} [classic]
\emph{organization of emergency repeat themselves on and on, and patients keep on dying}''.
\end{quote}

\end{itemize}


As can be seen from the above cases, a key cause of inefficiency in the interaction between users and
social organizations is often the fact that said interaction
is almost completely managed by the individual or by individual parts of 
the overall care system. \emph{The system does not work as a Whole, but rather as a collection of fragmented units}.
Moreover, the partial and outdated
knowledge that the individual and individual parts of the system have about the available social resources and their 
state often traslates in time-consuming and error-prone sequential queries (``pollings'').
We deem the major added value of FSO to be in the fact that it constructs a unitary Whole that takes responsibility
to assemble a coherent response to complex requests for service, thus avoiding the above inefficiencies.

\section{Scenarios}\label{s:scenarios}

The real-life cases in Sect.~\ref{s:failures} inspired the following scenario and its
reformulations.

Mary has just delivered her baby when she and her husband Joseph are told that the little one suffers from 
unexpected complications requiring specific support that is not available at the private clinic where they currently
are.

\subsection{Traditional Case} A call is immediately made to the emergency service (ES). On the phone, a verbal 
description of the situation and the immediate requirements is given. At the ES the case is tagged as life 
critical and immediate attention is taken. The ES operators have only partial knowledge; in particular they do 
know of a number of hospitals in the vicinity and of their generic specialty, but do not have an exact and 
up-to-date knowledge about each hospital's resources and their availability. An ordered list is made according to 
some criteria and calls are made to the hospitals in the list. The list includes three hospitals: St.~Thomas; 
City Polyclinic; and Regents. Through the calls, in a sequential fashion, it is realized that the second 
hospital, City Polyclinic, may be able to deal with the case at hand. An ambulance is dispatched and brings the little 
one and her parents to the chosen hospital. There, it is found out that the situation is more complex than
it had been depicted on the phone. Due to some misunderstanding and incomplete knowledge, it now becomes clear 
that the newborn baby is in need of special care that is not currently available at Polyclinic. Calls are 
made to the other two hospitals. One of them, Regents, could supply the sought special care, though they lack 
specific devices for extra-corporeal blood circulation, also needed for the case at hand and currently out of 
order at Regents. Another hospital with all the necessary is sought farther away. Had a complete knowledge been 
available, it would have transpired that St.~Thomas possessed the sought devices, and that mutualistic 
collaboration between Regents and St.~Thomas would have fulfilled all the prerequisites for dealing with 
the case.

\subsection{Infra-Community Cooperation}

A semantically annotated service request is published and reaches the coordinator of the nearest
SoC~\cite{DFCBD12,DF13c,DFC10}. The SoC has just a partial domain knowledge and raises an exception 
for lack of roles. The exception reaches a city SoC that includes the originating SoC and those of three 
hospitals (the above mentioned ones). Here, through live data and pub/sub, a more complete view of the available 
resources is available. It transpires that none of the members can fulfill the request on their own, but that a 
minimal set with all roles can be traced in two of the hospitals. A request to join is issued. The two SoC's 
``merge'' and become a SON (two members of the same SoC unite into a 
temporary new SoC). This is a trivial case of SON because enrolled members reside in the same community. The new 
SoC elects a coordinator (Regents') and enriches the request with new roles necessary to make it possible for the 
SoN to deal with the case at hand: an ambulance + driver must fetch the devices from St.~Thomas; at St.~Thomas, 
people to load/unload the device and technicians to operate the device must be available. Luckily agents can be 
assigned to the new necessary roles and the service protocol is started.

\subsection{Inter-community Cooperation}

A semantically annotated service request is published and reaches the coordinator of the nearest SoC.
The SoC has just a partial domain knowledge. Thanks to the fractal organization of the service, 
the SoC raises an exception for lack of roles. The exception reaches a city SoC that includes the originating SoC 
and those of three hospitals (the above mentioned ones). Here, through live data and pub/sub, a more complete 
view of the available resource is available. It transpires that none of the members can fulfill the request on 
their own. As a consequence, another exception is raised and reaches a regional SoC, including several city 
SoC's. It is now possible to identify that a minimal set with all roles can be traced in the hospitals of two 
city SoC's. A request to join is issued. The two SoC's ``merge'' and become a SON: two members of different SoC's 
unite into a temporary new SoC. The new SoC elects a coordinator (Regents') and enriches the request with new 
roles necessary to make it possible for the SON to deal with the case at hand: a helicopter + pilot must bring 
personnel from a hospital (Regent) to the other one. Missing roles produce a new exception, which is caught
by a civil defence SoC. Members of the latter include helicopters and helicopterists, two of which
are available. A mission is formulated and 
the service protocol is started.

\section{Model}\label{s:model}

\begin{figure}[t]
	\centerline{\includegraphics[width=0.60\textwidth]{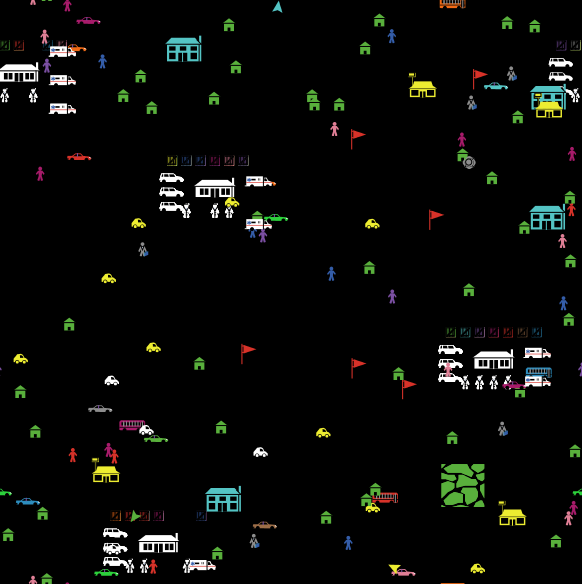}}
\caption{Netlogo screenshot of the simulation area with initialized agents.}\label{f:simarea}
\end{figure}

In this section we give a description of our simulation models and their functionality. For simulations we use the NETLogo 
environment~\cite{tisue_netlogo:simple_2004}, a tool that allows agent-based rapid prototyping. The simulation environment provides a simulation 
area that functions as a ``virtual world'' where the initialized agents perform their actions. These agents can move freely throughout the 
simulation 
area in discrete-time steps called ``ticks''. A visualization of the simulation area containing the initialized agents can be seen in 
Fig.~\ref{f:simarea}.

\begin{figure}[t]
	\centerline{\includegraphics[width=1.0\textwidth]{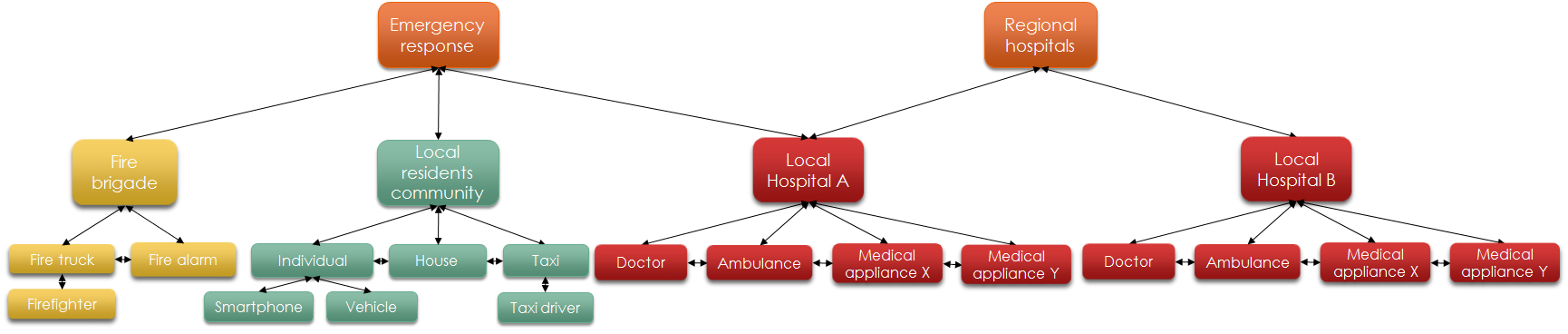}}
\caption{A representation of the hierarchy of SoCs used in our simulations.}\label{f:fsostruct}
\end{figure}

In our simulation models we have defined SoCs of different scale and levels of complexity. Figure~\ref{f:fsostruct} represents the hierarchical 
community structure used in experiments. The atomic components of the defined organization are individual SoCs, or iSoCs. An iSoC can be represented 
by a single individual (agent) and his/her personal belongings, for example an individual and his personal vehicle. In this case the individual 
represents the coordinator, or representative, of the iSoC while the vehicle is a member of the given iSoC. We define agents of diverse nature 
including Individuals, Doctors, Firefighters, and Taxi drivers. The emphasis here is on the individuals and their actions.

As the simulation starts each Individual decides to perform an action from a predefined set of six actions. The action can be a simple decision like 
``going to market'' or a more complex one, such as ``going to office'', in which the citizen has to be in office by a certain time. In the latter 
case, the Individual has to perform time, speed and distance calculations to find out the fastest way of reaching destination. By performing an 
activity Individuals implicitly play a Role within the Community. Another type of activity is ``going for a walk in park''. Whenever an activity is 
performed by an Individual, the representative of the corresponding Community is updated properly with the state of the agent and if an exception is 
triggered it will react accordingly. This allows the coordinating agent to manage the resources more intelligently. In the ``going for a walk in 
park'' 
case, if there is another Individual that has decided to perform the same activity, and is willing to do it together with someone, the coordinator 
establishes a mutualistic relationship between the two, so they perform the activity together.

The type of activities or events that we focus on in this paper are the ones where an Individual is in need of healthcare services. In such cases, 
the Individual takes the role of the patient. The severity of patient's health condition ranges from 1 to 10, where 1--3 represent a minor illness, 
meaning that the patient can pay a visit to local hospital by itself, whereas for 4--10 an ambulance is required to transport the Individual to one 
of the hospitals. Several hospitals are defined, each of them including a certain number of doctors, ambulances, and medical appliances. A doctor can 
treat all three minor illnesses and is an expert on three other diseases. In order to treat the patient the doctor uses corresponding medical 
appliances. The request for care service issued by a patient must be handled within a defined threshold, otherwise it is considered that the 
condition degrades and the patient dies. Based on the way the care request is handled we simulate three distinct scenarios, namely the Traditional 
Organization case, the FSO case, and a Perfect Oracle case. More information on these cases is given in what follows.


In the Traditional Organization case, the Individual in need picks a random hospital and issues a request. In cases when the condition of individual 
is larger than 3 the 
hospital responds by sending an ambulance to transport the patient. Once the patient reaches the hospital a doctor is assigned to treat the patient, 
and based on the patient's condition the appropriate medical appliances needed for treatment are allocated. It might happen that not all the 
resources required to treat the patient are available at the hospital. The missing resource could be, doctors that are experts on treating a 
particular condition, or specific medical appliances. In this case the patient's request is forwarded to another hospital. The procedure is repeated 
until either the patient receives the treatment or the request exceeds the predefined threshold---in which case the patient dies.

\begin{figure}[t]
  \begin{center}
    \subfigure[Patient waiting for ambulance.\label{f:waiting}]{\includegraphics[width=0.28\textwidth]{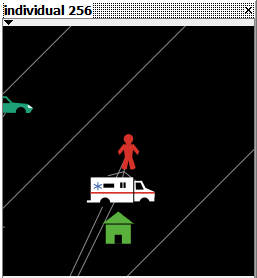}}
    \hspace{10pt}
    \subfigure[Ambulance transporting the patient.\label{f:transporting}]{\includegraphics[width=0.28\textwidth]{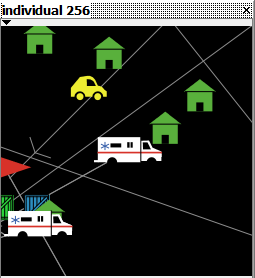}}
    \hspace{10pt}
    \subfigure[The arrival of patient to the hospital.\label{f:atHospital}]{\includegraphics[width=0.28\textwidth]{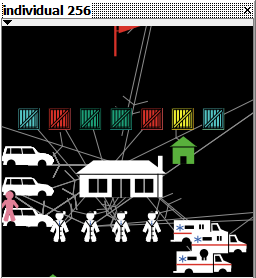}}
    \caption{Netlogo screenshot of phases that ``Individual 256'' goes through in case of ``healthcare service'' events.}
    \label{f:fsoevent}
  \end{center}
\end{figure}

The FSO approach treats the healthcare service events in a different manner. First, the individual in need of service raises an exception from its 
iSoC to the representative of the corresponding SoC (Local residents community). The representative or the coordinating agent handles the exception 
and checks for available resources in the current community. In case of a negative response, the exception is forwarded to the upper level 
representative, namely the ``Emergency response'' agent. Located at a higher hierarchical level, the ``Emergency response'' has broader knowledge 
regarding which agents might be able to deal with the situation at hand. After locating ``Local hospital A'' as a potential helping agent, the 
notification is directed towards it. ``Local Hospital A'' is well informed about its underlying agents and if all required resources are available it 
answers the exception by assigning a Doctor and possibly an Ambulance to deal with the case at hand. On the contrary, if ``Local Hospital A'' does 
not include all the available resources needed for handling the given situation, or in case the condition of the patient is complex and the hospital 
isn't able to cope with it, an exception is raised towards the higher-up community ``Regional hospitals''. ``Regional hospitals'' will be set in 
charge of finding another hospital in possession of the missing resources. With this setup the request will traverse the agents and spread throughout 
the layers of the FSO organization until the needed resources are found. This process is manifested by the establishment of a new SON, namely a 
temporary SoC consisting of resources ``taken'' from distinct SoCs, and it may be considered as the key factor behind FSO's effectiveness. A 
graphical representation of the phases a patient goes through before the start of its treatment is exemplified in Fig.~\ref{f:fsoevent}.

\begin{figure}[t]
  \begin{center}
    \subfigure[Infra-community cooperation.\label{f:logA}]{\includegraphics[width=0.90\textwidth]{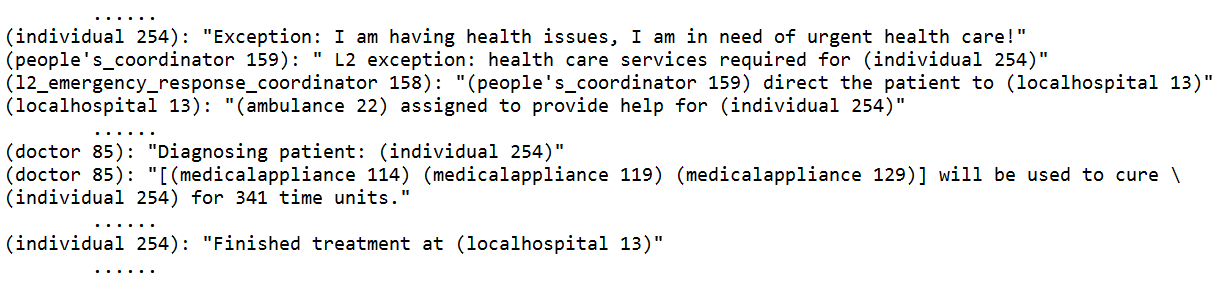}}
    
    \subfigure[Inter-community cooperation.\label{f:logB}]{\includegraphics[width=0.90\textwidth]{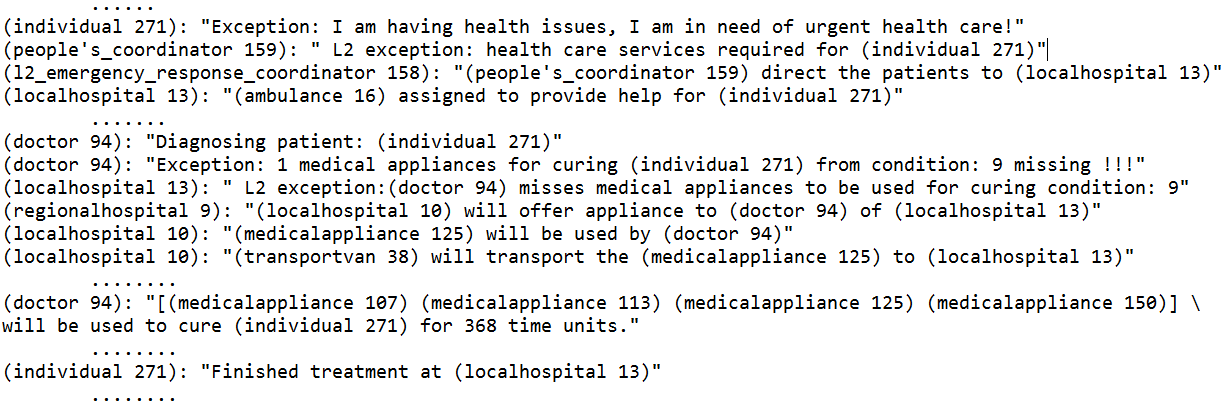}}\\[-10pt]
    \caption{An excerpt of the message flow between agents during event execution.}
    \label{f:logs}
  \end{center}
\end{figure}

Figure~\ref{f:logs} shows an extracted command log from the simulator representing the message flow between the agents. Figure~\ref{f:logA} shows a 
trivial case of SON in which the received exception is handled within the local hospital SoC---thus, without a need for additional resources---what 
we called in Sect.~\ref{s:scenarios} as infra-community cooperation. Figure~\ref{f:logB} represents a more complex case in which resources need to be 
shared smong two local hospitals---an example of inter-community cooperation.

In our simulation model we define three types of exceptions for local hospitals, namely:
\begin{itemize}
\item Lack of doctors capable of treating a particular type of patients disease.
\item No ambulances available to transport patients in severe condition.
\item Missing medical appliances needed to treat a patient.
\end{itemize}

By triggering exceptions whenever resources are missing within a particular hospital, an ``interaction channel'' between hospitals is established, 
thus allowing them to function as a greater organization---a ``greater Whole'' in which resources are shared in order to optimally respond to the 
case at hand.

Another model we have considered is the already mentioned Perfect Oracle. Here the Individuals are assumed to have perfect knowledge regarding 
the hospitals and services offered. However, with Perfect Oracle there is no resource sharing between the hospitals.
A possible scenario in this case is then one where the Individual in need reaches the hospital and once the doctor starts the treatment he or she 
finds out that there are no medical appliances available needed for treatment. In this case since the hospitals do not share their resources the 
patient's request is disregarded and it is directed to another hospital.


In what follows we model service requests as consisting of two components: the Querying Time (QT), representing the time from the moment an 
Individual is in need of treatment until the moment when all the needed resources are allocated, and the Treatment Time (TT), namely the time between 
start and end of treatment. Our experiments focus on QT as the organizational choice only affects that component. Said experiments are described in 
next section.

\section{Results}\label{s:results}

This section reports some preliminary results of the evaluation of the benefits of FSO with respect to Perfect Oracle (PO) and Traditional 
Organization (TO). As already mentioned, results are based on data collected from NETLogo simulation models. Simulation experiments lasted in all 
cases for 3000 ``ticks'' (simulation units) and involved the following set of community members: 4 Local hospitals, 15 Doctors, 8 Ambulances, 70 
Medical appliances, and a variable number of Individuals. The Doctors, Ambulances and Medical appliances are spread throughout the 4 hospitals 
according to the NETLogo random function, generated through a deterministic process. This means that using the same random seed the experiments can 
be reproduced. This was used in our experiments in order to have the same distribution of resources in all three models.
Individuals are involved in a number of activities at the end of which may become sick with a fixed probability of 0.09.
Average duration of activities is about 150 ticks.

\begin{figure}[h]
  \begin{center}
    \subfigure[Threshold 250.\label{f:threshold250}]{\includegraphics[width=0.35\textwidth]{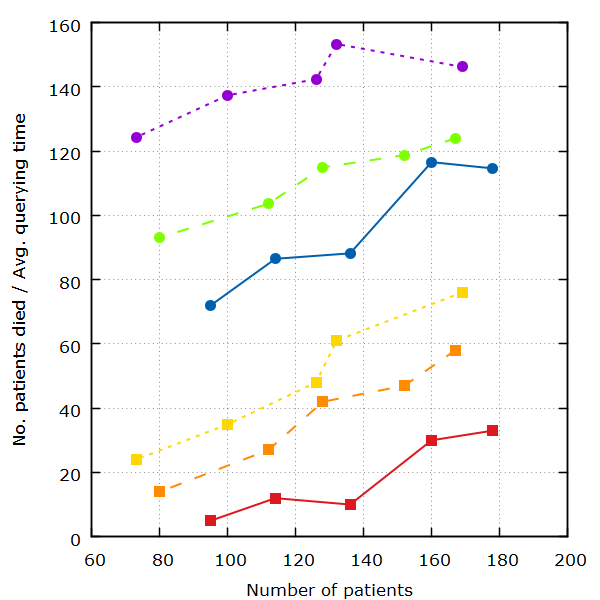}}
    \hspace{10pt}
    \subfigure[Threshold 200.\label{f:threshold200}]{\includegraphics[width=0.35\textwidth]{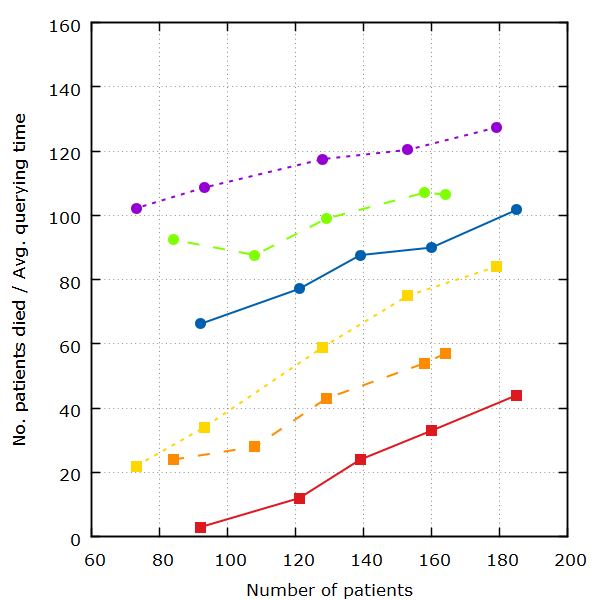}}
    \hspace{10pt}
    \subfigure[Threshold 150.\label{f:threshold150}]{\includegraphics[width=0.35\textwidth]{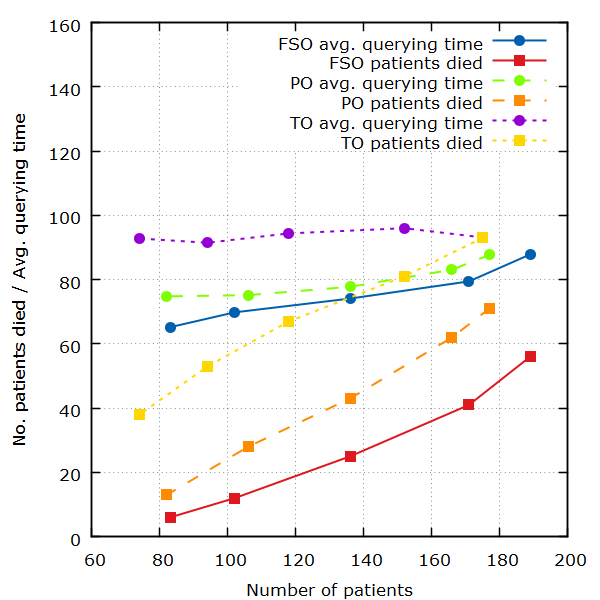}}
    \caption{Results from NETLogo simulations.}
    \label{f:experiments}
  \end{center}
\end{figure}

To evaluate the models we have used two main metrics, namely the average querying time and the number of patients which could not receive the 
treatment within the defined threshold, and thus died. We ran the experiments for each model with three different thresholds, as it can be observed 
in Fig.~\ref{f:experiments}. For each specified threshold the simulation was ran 5 times with a varying number of Individuals, starting from 60 to 
140. For FSO we have used an additional metric, SON, which describes the number of times a solution to the given situation is provided through 
inter-community cooperation. In a sense, this number also represents how much ``social energy'' is exploited by the communities. For TO
we also have measured the number of failures a patient faces until a matching hospital is found.

Figure~\ref{f:experiments} depicts the performance of all three models with the above described parameters. It can be observed that in all cases FSO 
performs better, followed by PO and then by TO. In the FSO case the number of patients that fail to receive the treatment is always lower---due to 
the sharing of resources between the hospitals---while for PO this number is higher because the patient knows at which hospital he or she should go. 
But if said hospital lacks, e.g., necessary medical appliances, it doesn't borrow them from other hospitals and redirects the patient instead. In the 
TO we have no interaction between hospitals neither. However, in this case the patient chooses the hospitals randomly, without any knowledge about 
the available doctors, ambulances or medical appliances. This poor level of information leads to delays in resource discovery, thus the number of 
patients that fail to receive treatment is the highest.

If we analyze the querying time results, again we see that FSO outperforms the other two models. This is due to the use of exceptions, which allows 
for fast inter-community communication and resource allocation. It can be observed that by decreasing the number of Individuals, the FSO improves 
drastically and the number of patients that fail to receive the treatment decreases two to three times with respect to the other cases.

The SON metric ranges from around 100 up to 220 depending on the number of Individuals. The higher the number of Individuals, the higher the SON 
metric. On the other hand, the number of service failures in the Traditional Organization ranges from around 50 up to 200. By increasing the number 
of Individuals the number of failures increases too.

\section{Conclusions}\label{s:end}

The time is \emph{now\/} to experimenting with new and more intelligent ways of organization and
management of resources. Turning around the current paradigms and having the organization tailor
an optimal response to the user's requests may provide mankind with one such way.
Evolving the traditional organizations---and learning how to do so without compromising the identity of the
intended services---is verily one of the greatest challenges our evolving humanity is to confront with.

Our answer to such challenge is given by FSO.
Working as a sort of ``cybernetic sociocracy'', FSO replaces
the plastic rigidity of traditional hierarchical-based organizations
with a dynamic infrastructure that allows cooperation among agents
of any scale and role. 
Intra- and inter-community collaboration is enabled by identifying chances
of mutualistic cooperation between members.
The evidence that we have started to collect and organize in this paper
makes us confident that
the new axioms of our FSO may make it possible to revert the concept of a
world ``too big'' and thus too difficult to manage into that of a
basin of seemingly unlimited \emph{social energy\/} produced by
a vast pool of intelligent mobile agents: we, the people.
Walt Kelly's famous reprimand---``We have met the enemy, and he is us''---which
summarizes the frustration produced by the \emph{old ways\/} applied to
an \emph{elderly world\/} would then be replaced with the more optimistic vision
of ``We have found the solution, and the solution is us''.

Many are the new challenges ahead.
Our plans include continuing our investigations in the theory and models of FSO as well as
in the process of gathering evidence to our claims both through simulation and
via real-life experimentations.
%
%


\bibliographystyle{splncs}

\end{document}